\documentclass[letterpaper,10pt,twocolumn]{revtex4}
\usepackage{amsmath,amssymb}
\usepackage{color}
\usepackage[dvips]{graphicx}
\definecolor{darkblue}{rgb}{0,0,0.7}
\usepackage[dvips, unicode, colorlinks, bookmarks=true, citecolor=darkblue, link
color=darkblue, urlcolor=darkblue]{hyperref}
\newcommand{\hence}{\ \Rightarrow\ }
\newcommand{\svector}[2]{\begin{pmatrix}#1\\#2\end{pmatrix}}
\newcommand{\smatrix}[4]{\begin{pmatrix}#1 & #2\\#3 & #4\end{pmatrix}}

\begin{document}

\title{Pass-through Mach-Zehnder topologies for macroscopic quantum measurements}

\author{F.Ya.Khalili}

\affiliation{Moscow State University}


\begin{abstract}

Several relatively small-scale experimental setups aimed on prototyping of future laser gravitational-wave detectors and testing of new methods of quantum measurements with macroscopic mechanical objects, are under development now. In these devices, not devoted directly to the gravitational-wave detection, Mach-Zehnder interferometer with pass-through Fabry-Perot cavities in the arms can be used instead of the standard Michelson/Fabry-Perot one. The advantage of this topology is that it does not contain high-reflectivity end mirrors with multilayer coatings, which Brownian noise could constitute the major part of the noise budget of the Michelson/Fabry-Perot interferometers.

We consider here two variants of this topology: the ``ordinary'' position meter scheme, and a new variant of the quantum speed meter.

\end{abstract}

\maketitle



\section{Introduction}

Sensitivity of modern large-scale laser gravitational-wave detectors \cite{Waldman2006, Acernese2006, Hild2006, Ando2005} is limited by two groups of noise sources. The first one comprises so-called ``classical'', or ``technical'' noises of different origin --- thermal, seismic, {\it etc}, which, in principle, can be reduced by cooling, using better materials, more sophisticated seismic isolation and so on. The second one originates from the  quantum fluctuations of the light phase and amplitude inside the interferometers. These noise sources --- the shot noise and the radiation pressure noise, correspondingly --- obey the fundamental limitation imposed by Heisenberg's uncertainty relation \cite{Caves1981}. ``Naive'' optimization of the quantum noise, which simply makes the shot noise and the radiation pressure noise equal to each other, imposes a so-called Standard Quantum Limit \cite{92BookBrKh} of the sensitivity:
\begin{equation}\label{S_SQL} 
  S_{\rm SQL} = \frac{2\hbar}{M\Omega^2} \,,
\end{equation} 
where $S_{\rm SQL}$ is the equivalent position noise spectral density \footnote{Single-sided normalization of the spectral density standard for the gravitational-wave community is used in this paper}, $M$ is the test mass and $\Omega$ is the observation frequency. This limitation can be evaded using more sophisticated measurement schemes (see, {\it e.g.} \cite{90a1BrKh, 99a1BrKh, Buonanno2001, 02a1KiLeMaThVy, Chen2002, Corbitt2004-3}), which, probably, will be implemented in the third generation gravitational-wave detectors \cite{09a1ChDaKhMu, ETsite}.

The second-generation gravitational-wave detectors, which are under development now \cite{Harry2010, AdvLIGOsite, AdvVIRGOsite, LCGTsite}, are typically described as ``quantum-noise limited'' ones. However, in the best sensitivity band (around $100\,{\rm Hz}$), the noise budget will be dominated by one of the ``classical'' noises, namely, by the thermal fluctuations of thickness of the multilayer dielectric coatings of the interferometers mirrors \cite{Harry2002, Penn2003, 03a1BrVy, Fejer2004, Harry2006, 08a1Go}.

\begin{figure}
  \includegraphics{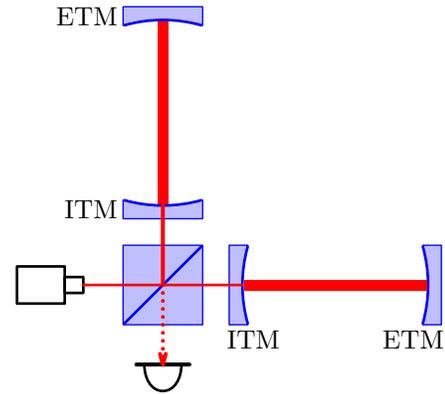}
  \caption{Simplified scheme of the Michelson interferometer topology of laser gravitational-wave detectors. ITM: input test masses; ETM: end test masses.}\label{fig:michelson}
\end{figure} 

All contemporary and planned large-scale laser gravitational-wave detectors have the same Michelson/Fabry-Perot topology, shown in Fig.\,\ref{fig:michelson}. Two arms here provide balanced scheme which suppresses the laser technical noise, and the arms orthogonal placement is dictated by the spatial structure of gravitational waves. In this topology, transmittance of the end mirrors (ETM) have to be as small as possible, while the input mirrors (ITM) have to have some moderate transmittance which depends on the required bandwidth of the arm cavities (the typical values are $T^2_{\rm ETM}\sim10^{-5}$ and $T^2_{\rm ITM}\lesssim10^{-2}$, respectively). Therefore, the end mirrors have to have at least $2\,\text{-}\,3$ times more coating layers, than the input ones. Taking into account that the coating thermal noise spectral density with good precision is proportional to the coating layers number, this means that it is the end mirrors produce the major part of this noise.

Due to this reason, it was proposed several years ago to replace the end mirrors by coatingless corner reflectors \cite{04a1BrVy} or by two-mirror reflectors \cite{05a1Kh}. Later analysis  showed, however, that the corner reflectors introduce significant optical losses \cite{05a1Ta}. 

In addition to full-scale laser gravitational-wave detectors, several smaller scale interferometers not devoted directly to the gravitational-wave detection are under development now \cite{Gossler_CQG_2010, 10m_site, Zhao2006, Gingin_site, Miyoki_JP_2010_203_012075, CLIOsite}, with the goal of testing different aspects of future gravitational-wave detectors, in particular, new methods of quantum measurements with macroscopic mechanical objects. These devices have the same Michelson/Fabry-Perot topology of Fig.\,\ref{fig:michelson}. However, in this case, another topology can be used, which does not contain the high-reflectivity end mirrors at all: the well known Mach-Zehnder interferometer.

Similar to the Michelson/Fabry-Perot topology, Fabry-Perot cavities can be placed inside the Mach-Zehnder interferometer arms in order to increase the optomechanical coupling, see Fig.\,\ref{fig:mach_zehnder}. In this case, both input and end mirrors of the Fabry-Perot cavities must have the same moderate reflectivities created by means of a several reflective layers, instead of tens in the high-reflectivity end mirrors, and, therefore, proportionally smaller coating thermal noise.

It can be noted, that in principle, Mach-Zehnder/Fabry-Perot topology can be used in the large-scale gravitational-wave detectors case too. However, in this case the second beamsplitter has to recombine beams from the end mirrors separated by the several kilometers distance. It is unclear, whether the sensitivity gain provided by this topology justify the corresponding technological hassles.

%

The evident difference of the Mach-Zehnder/Fabry-Perot topology from the Michelson/Fabry-Perot one is that it has two output ports, with outgoing fields in both of them carrying information about the mirrors motion. In order to achieve a quantum noise limited sensitivity, all this information has to be captured. Therefore, both dark ports have to be equipped with photodetectors.

In this topology, all proposed methods of overcoming the SQL can be used too. But its two output ports provide an additional flexibility in implementation of these methods. In particular, it can be converted into the {\it quantum speed meter} \cite{90a1BrKh}, by adding two small mirrors into the centers of each of the Fabry-Perot cavities, see Fig.\,\ref{fig:speedmeter}. The main advantage of the speed measurement is that it allows to overcome the SQL in broad band simply by using homodyne detector with frequency-independent homodyne angle. In the ``ordinary'' position measurement case, sophisticated schemes which provide frequency-independent homodyne angle or frequency-independent squeezing are required for this purposes \cite{02a1KiLeMaThVy}.

The topology of Fig.\,\ref{fig:speedmeter} is simpler and more suitable for implementation in relatively small-scale prototype devices, than other proposed ``flavours'' of the quantum speed meter: Doppler meter \cite{96a1KhLe}, ``sloshing cavities'' scheme \cite{00a1BrGoKhTh, Purdue2002}, and Sagnaq interferometer \cite{Chen2002, 04a1Da}. However, instead of pure velocity information, it provides information about sophisticated mix of all six mirrors positions and velocities \cite{Chen_private}. We show here, that all the irrelevant information can be suppressed by using sufficiently small ratio of the central/end mirrors masses and proper choice of the homodyne angles of the two detectors.

The paper is organized as follows. In Sec.\,\ref{sec:pos}, we calculate quantum noise of the Mach-Zehnder/Fabry-Perot interferometer and show equivalence of this topology to the Michelson/Fabry-Perot one with ideally reflective end mirrors. In Sec.\,\ref{sec:speed}, we consider an implementation of the {\it quantum speed meter} scheme based on the Mach-Zehnder/Fabry-Perot topology. In Sec.\,\ref{sec:conclusion} we resume the obtained results.

The main notations used in this paper are listed in Table\,\ref{tab:notations}. The Appendices contain all calculations which are not necessary for understanding the main results of this paper.

\begin{table}[t]
  \begin{tabular}{|c|l|}
    \hline
      Quantity    & Description \\
    \hline
      $\Omega$    & Observation (sideband) frequency \\
      $c$         & Speed of light \\
      $M$         & Arm cavities mirrors mass \\
      $L$         & Arm cavities length \\
      $\omega_o$  & Arm cavities eigen frequency \\
      $\gamma$    & Arm cavities half-bandwidth \\
      $I_c$       & Power circulating in each of the arm cavities \\
      $\eta$      & Unified quantum efficiency \\
      $m$         & Central mirrors mass in the speed meter scheme \\
    \hline
  \end{tabular}
  \caption{Main notations used in this paper.}\label{tab:notations}
\end{table}

\section{Two-ports position meter}\label{sec:pos}

\begin{figure}
  \includegraphics{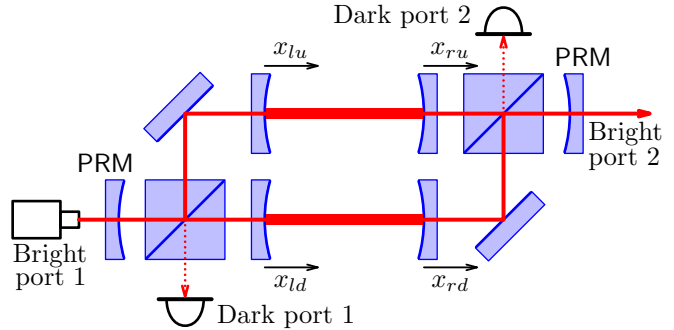}
  \caption{The scheme of Mach-Zehnder/Fabry-Perot interferometer.}\label{fig:mach_zehnder} 
\end{figure} 

Consider the Mach-Zehnder/Fabry-Perot interferometer shown in Fig.\,\ref{fig:mach_zehnder}. In this paper, we suppose that the arm cavities here are tuned in resonance, and the input and the end mirrors have the same transmittance $T^2$. In this case, most of the optical power which enters the interferometer through one of the bright ports (left in the picture) leaves it through the second (right) bright port, but some small fraction, proportional to the differential displacement of the mirrors:
\begin{equation}
  x = \frac{x_r-x_l}{2} \,,
\end{equation}
where
\begin{align}
  & x_r = \frac{x_{ld}-x_{lu}}{2} \,, & x_r = \frac{x_{rd}-x_{ru}}{2}
\end{align}
(subscripts $l$ and $r$ stand for ``left'' and ``right'' mirrors, and $u$ and $d$ --- for ``down'' and ``up'' cavities), goes into the dark ports, where it can be detected. We assume here that both dark ports are equipped by homodyne detectors with the homodyne angles $\phi_{1,2}$.

Two additional power recycling mirrors ({\sf PRM}), shown in Fig.\,\ref{fig:mach_zehnder}, in principle, are not necessary for the scheme function. However, they allows to increase the optical power circulating in the Fabry-Perot cavities (for the same laser power and for the same finesse of the cavities).

\begin{figure}
  \includegraphics{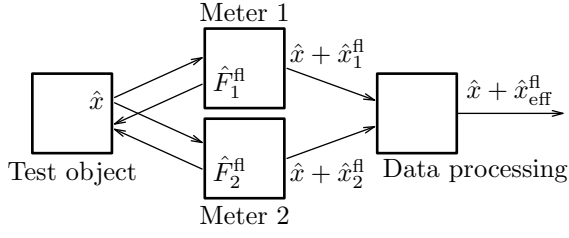}
  \caption{Generalized scheme of the dual linear quantum meter.}\label{fig:double}
\end{figure} 

It is shown in Appendix \ref{app:pm} that this scheme is equivalent to two independent quantum position meters which simultaneously measure position $x$ of the same mechanical degree of freedom, see Fig.\,\ref{fig:double}. Output signals of these meters can be represented as sums of the actual position $x$ of the test object and the corresponding measurement noises $\hat{x}^{\rm fl}_{1,2}$ (originated from the light shot noise):
\begin{equation}\label{pm_x_12} 
  \tilde x_{1,2} = \hat{x} + \hat{x}^{\rm fl}_{1,2} \,.
\end{equation}

Both meters also apply back-action noises $\hat{F}^{\rm fl}_{1,2}$ (fluctuations of the radiation-pressure forces) on the test object. Therefore, in spectral domain, the test object position can be represented as follows:
\begin{equation}
  \hat{x}(\Omega) = \hat{x}_0(\Omega)
    + \chi(\Omega)[\hat{F}_1^{\rm fl}(\Omega) + \hat{F}_2^{\rm fl}(\Omega)] ,
\end{equation}
where $\hat{x}_0(\Omega)$ of the intrinsic motion of the test object, that is the one independent on the meter noises (in particular, it can be the response on some external signal force which has to be detected), and $\chi(\Omega)$ is the susceptibility function of the test object (for example, $\chi(\Omega) = -1/(m\Omega^2)$ for a free mass $m$).

Spectral densities of these noises: $S_{1,2}^x$, $S_{1,2}^F$, and the corresponding cross-correlation spectral densities $S_{1,2}^{xF}$ are calculated in Appendix \ref{app:pm}, see Eqs.\,\eqref{pm_SxSF}. They have exactly the same form as in the quantum noise spectral densities of the ordinary Michelson/Fabry-Perot interferometer, but with halved circulating power (note that here two independent effective meters share the same optical power). In particular, these spectral densities satisfy the following uncertainty relation \cite{92BookBrKh}:
\begin{equation}\label{pm_SxSF_prod} 
  S_{1,2}^xS_{1,2}^F - |S_{1,2}^{xF}|^2 \ge \hbar^2 \,.
\end{equation}
with both sides being equal in the ideal case of no optical loss ($\eta = 1$).

Two output signals \eqref{pm_x_12} should be combined by the optimal data processing scheme, giving:
\begin{multline}\label{x12_to_y}
  \tilde x_{\rm sum}(\Omega)
  = \xi(\Omega)\tilde x_1(\Omega) + [1-\xi(\Omega)]\tilde x_2(\Omega) \\
  = \hat{x}_0(\Omega) + \hat{x}^{\rm sum}(\Omega) \,,
\end{multline}
where $\xi(\Omega)$ is the weight function which has to be optimized and
\begin{multline}
  \hat{x}^{\rm sum}(\Omega)
  = \xi(\Omega)\hat{x}^{\rm fl}_1(\Omega) + [1-\xi(\Omega)]\hat{x}^{\rm fl}_2(\Omega) \\
  + \chi(\Omega)[\hat{F}_1^{\rm fl}(\Omega) + \hat{F}_2^{\rm fl}(\Omega)]
\end{multline} 
is the sum noise. It should be emphasized that transformation \eqref{x12_to_y} is a software operation which can be performed {\it post factum}. Therefore, no limitations are applied to the function $\xi(\Omega)$; in particular, this filtering can be acausal one.

Spectral density of the $\hat{x}^{\rm sum}$ noise is equal to
\begin{multline}
  S_{\rm sum} = |\xi(\Omega)|^2S_1^x + |1-\xi(\Omega)|^2S_2^x \\
  + 2\Re\Bigl(
      \chi(\Omega)\left\{\xi^*(\Omega)S_1^{xF} + [1-\xi^*(\Omega)]S_2^{xF}\right\}
    \Bigr) \\
  + |\chi(\Omega)|^2(S_1^F + S_2^F) \,.
\end{multline}
The minimum of this spectral density in $\xi$ is equal to
\begin{equation}\label{PTH_gen_Ssum}
  S^x = S_x^{\rm eff} + 2\Re[\chi(\Omega)S_{xF}^{\rm eff}]
    + |\chi(\Omega)|^2S_F^{\rm eff} \,,
\end{equation}
where
\begin{subequations}\label{PTH_S_eff}
  \begin{gather}
    S^x_{\rm eff} = \frac{S_1^xS_2^x}{S_1^x + S_2^x} \,, \\
    S^F_{\rm eff} = S_1^F + S_2^F - \frac{|S_1^{xF} - S_2^{xF}|^2}{S_1^x + S_2^x}\,,\\
    S^{xF}_{\rm eff} = \frac{S_1^xS_2^{xF} + S_2^xS_1^{xF}}{S_1^x + S_2^x}
  \end{gather}
\end{subequations}
are, correspondingly, spectral densities of the effective measurement noise, the effective back action noise, and the cross-correlation spectral density for this scheme.  It is easy to show, if both meters are ideal quantum noise limited ones, then the same is valid for the effective combined meter.

Consider the symmetric particular case of the homodyne angles equal to each other:
\begin{equation}\label{pm_eq_phis} 
  \phi_1=\phi_2=\phi \,.
\end{equation} 
If follows from Eqs.\,\eqref{pm_SxSF}, that it corresponds to two identical meters:
\begin{subequations}
  \begin{gather}
    S_1^x = S_2^x = 2S^x_{\rm eff} , \\
    S_1^F = S_2^F = \frac{S^F_{\rm eff}}{2} , \\
    S_1^{xF} = S_2^{xF} = S^{xF}_{\rm eff} \,.
  \end{gather}
\end{subequations}
where the effective noise spectral densities are equal to
\begin{subequations}
  \begin{gather}
    S^x_{\rm eff} = \frac{\hbar cL}{16\omega_oI_c\gamma\eta}\,
      \frac{\gamma^2+\Omega^2}{\cos^2\phi} \,, \\
    S^F_{\rm eff} = \frac{16\hbar\omega_oI_c\gamma}{cL(\gamma^2+\Omega^2)} \,, \\
    S^{xF}_{\rm eff} = \hbar\tan\phi \,.
  \end{gather}
\end{subequations}
These spectral densities are exactly equal to the ones of the Michelson/Fabry-Perot interferometer with the same parameters, in particular, with the same unified quantum efficiency $\eta$ and the same half-bandwidth $\gamma$. But while in the Michelson/Fabry-Perot case the inefficiency $1-\eta$ is created by the photodiods quantum inefficiency, losses in the optical elements, and {\it transmittance of the end mirrors}, in the Mach-Zehnder/Fabry-Perot case the last component is absent. Therefore, the Mach-Zehnder/Fabry-Perot interferometer with symmetric homodyne angles \eqref{pm_eq_phis} is equivalent to the Michelson/Fabry-Perot interferometer {\it with ideally reflective end mirrors}.

It should be noted, however, that in order to keep the same value of the half-bandwidth $\gamma$, the arm cavities mirrors transmittances in the Mach-Zehnder/Fabry-Perot interferometer must be twice as small as the ones of Michelson/Fabry-Perot, because in the former case, both input and end mirrors transmittances introduce equal parts into $\gamma$.

Consider, for example, the Michelson/Fabry-Perot interferometer with the input and end mirrors power transmittances equal to $T^2_{\rm ITM}=0.1$ and $T^2_{\rm ETM}=10^{-5}$. These values can be considered as typical ones for the smaller-scale prototype interferometer, similar to the AEI 10-m interferometer \cite{Gossler_CQG_2010, 10m_site}, and they can be created using $N_{\rm ITM}=5$ and $N_{\rm ETM}=18$ layers of ${\rm Ta}_2{\rm O}_5$, respectively (see, {\it e.g.}, Eq.\,(2) of \cite{Somiya2008}), giving total number of ${\rm Ta}_2{\rm O}_5$ layers for each Fabry-Perot cavity equal to 23.

At the same time, in order to obtain the same value of the bandwidth in the Mach-Zehnder/Fabry-Perot case, all arm cavities mirrors have to have transmittance equal to $T^2=0.05$, which corresponds to 6 layers for each mirror and 12 layers total. This almost twice as small number of the coating layers translates to the proportional decrease of the coating noise spectral density.

\section{Two-port speed meter}\label{sec:speed}

\begin{figure}
  \includegraphics{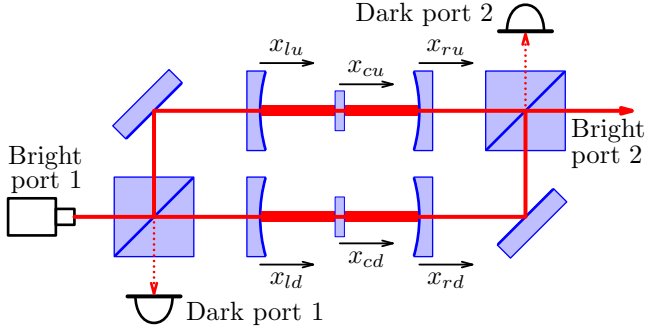}
  \caption{The scheme of the Mach-Zehnder/Fabry-Perot quantum speed meter.}\label{fig:speedmeter}
\end{figure} 

The quantum speed meter version of the Mach-Zehnder/Fabry-Perot interferometer is shown in Fig.\,\ref{fig:speedmeter}. It differs from the previous scheme by two additional small mirrors located in the centers of the Fabry-Perot cavities. Both surfaces of these mirrors should have reflective coatings with the transmittances equal to one of the Fabry-Perot cavities main mirrors, and the short Fabry-Perot etalons formed thus inside the additional mirrors should be tuned in anti-resonance: the optical length shall be equal to
\begin{equation}\label{anti-res} 
  d = \frac{\lambda(n+1/2)}{2} \,,
\end{equation}  
where $n$ is an integer.

\begin{figure}
  \includegraphics{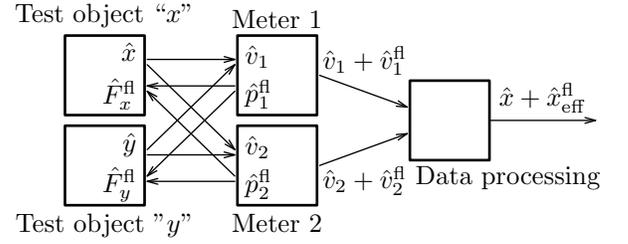}
  \caption{Generalized scheme of the Mach-Zehnder/Fabry-Perot quantum speed meter.}\label{fig:sp_double}
\end{figure} 

Quantum noise of this configuration is calculated in Appendix \,\ref{app:sm}. It is shown, that it is equivalent to the generalized scheme shown in Fig.\,\ref{fig:sp_double}. Here two independent meters (which correspond to two dark ports of Fig.\,\ref{fig:speedmeter}) interact simultaneously with two mechanical degrees of freedoms, which correspond to the following mechanical modes:
\begin{align}
  & x = x_c - \frac{x_r+x_l}{2} \,, & & y = \frac{x_r-x_l}{2} \,,
\end{align}
where
\begin{align}
  & x_r = \frac{x_{ld}-x_{lu}}{2} \,, & x_c = \frac{x_{cd}-x_{cu}}{2} \,, &
  & x_r = \frac{x_{rd}-x_{ru}}{2} 
\end{align}
(subscripts $l$, $c$, $r$ stand for ``left'', ``central'', ``right'' mirrors, and $u$ and $d$ --- for ``down'' and ``up'' cavities). Effective masses of these degrees of freedom are equal to
\begin{equation}\label{sm_mu} 
  \mu = \left(\frac{1}{m}+\frac{1}{2M}\right)^{-1}
\end{equation}
and $2M$, correspondingly.

In spectral picture, output signals of these two meters can be represented as follows:
\begin{equation}\label{sm_v12}
  v_{1,2}(\Omega) = X_{1,2}(\Omega)x(\Omega) + Y_{1,2}(\Omega)y(\Omega)
    + v_{1,2}^{\rm fl}(\Omega) \,,
\end{equation} 
where
\begin{subequations}\label{sm_XY} 
  \begin{align}
    & X_1(\Omega) = \gamma-i\Omega \,,  & & Y_1 = -i\Omega \,, \\
    & X_2(\Omega) = -i\Omega \,,        & & Y_2 = -(\gamma-i\Omega) \,,
  \end{align}
\end{subequations}
and $v_{1,2}^{{\rm fl}}$ are the measurement noises. Correspondingly, both meters together create two fluctuational forces acting on the test objects:
\begin{subequations}\label{sm_F_fl_xy}
  \begin{gather}
    F_x^{\rm fl}(\Omega) = X_1^*(\Omega)\hat{p}_1^{\rm fl}(\Omega)
      + X_2^*(\Omega)\hat{p}_2^{\rm fl}(\Omega) \,, \\
    F_y^{\rm fl}(\Omega) = Y_1^*(\Omega)\hat{p}_1^{\rm fl}(\Omega)
      + Y_2^*(\Omega)\hat{p}_2^{\rm fl}(\Omega) \,,
  \end{gather}
\end{subequations}
where $\hat{p}^{\rm fl}_{1,2}$ are the back-action noises. Spectral densities $S^v_{1,2}$ and $S^p_{1,2}$ of the noises $v_{1,2}^{\rm fl}$ and $p_{1,2}^{\rm fl}$, together with the cross-correlation spectral densities $S^{vp}_{1,2}$ [see Eqs.\,\eqref{sm_S_vp}] satisfy the standard uncertainty relation similar to \eqref{pm_SxSF_prod},
\begin{equation}
  S^v_{1,2}S^p_{1,2} - |S^{vp}_{1,2}|^2 \ge \hbar^2 \,,
\end{equation} 
again with the exact equality in the no-losses case of $\eta=1$.

It follows from Eq.\,\eqref{sm_v12}, that the first meter provides information about the position of the ``$x$'' degree of freedom and velocity of the ``$y$'' degree of freedom, and the second meter --- about the velocity of the ``$x$'' degree of freedom and position of the ``$y$'' degree of freedom. In order to implement the quantum speed meter regime, the position information has to be suppressed. It can be done in the following way.

First, suppose that the end mirrors are very heavy compared to the central ones:
\begin{equation}
  \frac{\mu}{M} \approx \frac{m}{M} \to 0 \,.
\end{equation} 
(the general case of finite masses ratio will be considered below). In this case, the end mirrors can be treated as fixed ones, $y=0$, and the measurement scheme reduces to the one considered in the previous section: two meters and one test object, see Fig.\,\ref{fig:double}, with the noise spectral densities equal to:
\begin{align}\label{sm_no_M} 
  & S_{1,2}^x = \frac{S_{1,2}^v}{|X_{1,2}|^2} \,, &
  & S_{1,2}^F = |X_{1,2}|^2S_{1,2}^p \,, &
  & S_{1,2}^{xF} = S_{1,2}^{vp} \,.
\end{align}
Thus, only one source of undesirable information remains: the first meter providing position information about the ``$x$'' degree of freedom.

Then, suppose that the first homodyne detector measures the amplitude quadrature of the outgoing light, which corresponds to the homodyne angle
\begin{equation}\label{sm_phi_1} 
  \phi_1 = \frac{\pi}{2} \,.
\end{equation} 
This means that the first meter does not ``see'' the input signal, but instead ``sees'' its own back action force. Its noise spectral densities in this case have the following structure:
\begin{align}\label{virt_mirr}
  & S_1^{xF} \to \infty &
  & S_1^x = \frac{\hbar^2 + |S_1^{xF}|^2}{\eta S_1^F} \to \infty \,.
\end{align}
Substitution of Eqs.\,(\ref{sm_no_M}, \ref{virt_mirr}) into Eqs.\,\eqref{PTH_S_eff} gives the following effective spectral densities:
\begin{subequations}\label{sm_S_eff} 
  \begin{gather}
    S^x_{\rm eff} = S^x_2 \,, \\
    S^F_{\rm eff} = S^F_2 + (1-\eta)S^F_1 \,, \label{sm_nonsymm(b)} \\
    S^{xF}_{\rm eff} = S^{xF}_2 \,.
  \end{gather}
\end{subequations}
They are almost exactly equal to the spectral densities of the second meter alone. The only difference is the small term in Eq.\,\eqref{sm_nonsymm(b)} proportional to the quantum inefficiency $1-\eta$. Therefore, the first meter with its undesirable (position measurement) noises structure is virtually excluded from the measurement process: due to the homodyne angle setting \eqref{sm_phi_1}, it does not provide any information about the test object motion, and the optimal data processing procedure almost completely removes the results of the first meter back action from the output signal (while the real motion of the test object, of course, is perturbed by both meters). Thus, this meter plays the role of a ``virtual mirror'' which effectively closes the corresponding output port of the interferometer.

As a result, only one information channel remains: the second meter which provides information about the velocity of the central mirrors differential motion. The sum noise spectral density of this measurement is equal to [see Eqs.\,(\ref{sm_S_vp}, \ref{sm_no_M}, and \ref{sm_S_eff})]:
\begin{multline}\label{sm_S_sum0}
  S_{\rm sum} = S^x_{\rm eff} - \frac{2S^{xF}_{\rm eff}}{\mu\Omega^2}
    + \frac{S^F_{\rm eff}}{\mu^2\Omega^4} \\
  = \frac{\hbar}{\mu\Omega^2}\biggl\{
      \frac{1}{\mathcal{K}\eta\cos^2\phi_2} - 2\tan\phi_2 \\
      + \mathcal{K}\left[1 + \frac{(1-\eta)(\gamma^2+\Omega^2)}{\Omega^2}\right]
    \biggr\} ,
\end{multline}
where
\begin{align}\label{sm_varK}
  & \mathcal{K} = \dfrac{\mathcal{K}_0}{1+4\Omega^4/\gamma^4} \,, &
  & \mathcal{K}_0 = \dfrac{32\omega_oI_c}{\mu cL\gamma^3} \,.
\end{align}

This spectral density has typical speed meter frequency dependence, with the optomechanical coupling factor $\mathcal{K}$ being asymptotically constant and reaching its maximum $\mathcal{K}_0$ at $\Omega\to0$ (compare, {\it e.g.}, with Eqs.\,(20, 21) of Ref.\,\cite{09a1ChDaKhMu}). Therefore, the following homodyne angle:
\begin{equation}\label{sm_phi_0} 
  \phi_2 = \arctan(\mathcal{K}_0\eta)
\end{equation} 
provides broad-band low-frequency optimization of the spectral density \eqref{sm_S_sum0}, which gives:
\begin{multline}\label{sm_S_sum} 
  S_{\rm sum} = \frac{\hbar}{\mu\Omega^2}\biggl\{
    \frac{1}{\mathcal{K}\eta} \\ 
    + \mathcal{K}\eta\left[
        \frac{16\Omega^8}{\gamma^8}
        + \frac{1-\eta}{\eta}\left(2+\frac{\gamma^2}{\Omega^2}\right)
      \right]
  \biggr\} .
\end{multline} 
It is easy to see that in the absence of the optical losses, $1-\eta\to0$, the residual back action noise (proportional to $\mathcal{K}$) can be very small at low frequencies, $\Omega<\gamma$. However, similar to other speed meter implementations \cite{09a1ChDaKhMu}, the additional term introduced in Eq.\,\eqref{sm_S_sum} by the optical losses, diverges if $\Omega\to0$ and therefore seriously limits the sensitivity.

The sum noise spectral density for the general case of arbitrary masses ratio $m/M$ is calculated in Appendix \ref{app:sm_S_sum}, see Eqs.\,(\ref{sm_S_sum_k}, \ref{sm_S_sum_k_opt}). In this case, the perturbation of the ``$y$'' degree of freedom creates additional terms in the sum noise spectral density which also diverge at $\Omega\to0$. However, $m/M$ can be easily made much smaller than the quantum inefficiency $1-\eta$. Therefore, the  sensitivity degradation due to the end mirrors motion perturbation can be made negligibly small in comparison with the one caused by the optical losses.

\begin{figure}
  \includegraphics{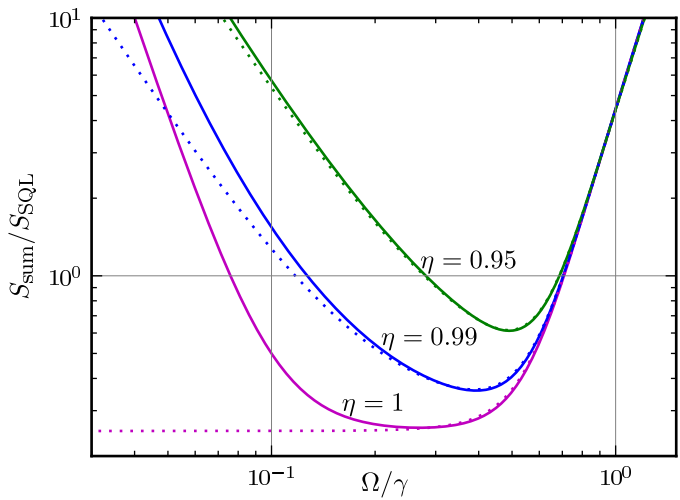}
  \caption{Plots of $\dfrac{\mu\Omega^2}{\hbar}\,S_{\rm sum}$ [see Eq.\,\eqref{sm_S_sum}] for $m/M=0.01$ (solid lines) and $m/M\to0$ (dots). For all plots, $\mathcal{K}_0=2$.}\label{fig:plots}
\end{figure} 

In Fig.\,\ref{fig:plots}, spectral density \eqref{sm_S_sum} normalized by the SQL spectral density $\hbar/(\mu\Omega^2)$ is plotted for $m/M=0.01$ and for the asymptotic case $m/M\to0$, for the following three values of the unified quantum efficiency: $\eta=0.95$ (moderately optimistic), $\eta=0.99$ (very optimistic), and $\eta=1$ (no losses at all). In all these cases, the value of $\mathcal{K}_0=2$ is used. If, for example, $2\pi c/\omega_o = 1.064\,\mu\mathrm{m}$, $m=10\,{\rm g}\hence M=1\,{\rm kg}$, $L=5\,{\rm m}$ (total arm length $2L=10\,{\rm m}$), $T^2=10^{-3}$ (these values are close to ones planned for Hannover 10-m prototype experiment \cite{Gossler_CQG_2010}), then this $\mathcal{K}_0$ corresponds to $I_c\approx15\,{\rm kW}$.

It is easy to see, that influence of the ``$y$'' degree of freedom is barely noticeable if $\eta=0.95$, small if $\eta=0.99$ and is significant only in the ideal lossless case $\eta=1$.

It have to be noted that the reduced size of the central mirror requires that the beam radius on this mirrors also have to be reduced proportionally, in order to keep diffraction losses at the same
level. The reduced beam radius means increased influence of Brownian and thermoelastic noises of the central mirror coating and bulk \cite{03a1BrVy}. However, estimates made in \cite{06a1Kh} show that this effect is more than compensated by the increased Standard Quantum Limit \eqref{S_SQL} of the light central mirrors.

Another potential noise source is thermo-refractive noise in the central mirrors bulk \cite{00a1BrGoVy}. However, due to the condition \eqref{anti-res}, the optical power penetrating into the central mirrors will be as small as the power inside the Fabry-Perot cavities end mirrors and  the beamsplitters. Therefore, due to the longer light path inside, it is these elements will contribute the major part of the total thermo-refractive noise.

\section{Conclusion}\label{sec:conclusion}

The brief estimate made above shows that in the Mach-Zehnder/Fabry-Perot interferometer, the coating thermal noise spectral density can be about twice as small as in the similar Michelson/Fabry-Perot. Taking into account, that this gain is cumulative with the other proposed methods of reducing the coating thermal noise: in particular, broadened laser beams \cite{Bondaresku2006}, better coating materials \cite{Harry2007}, coating structure optimization \cite{Kimble_PRL_101_260602_2008, Villar2010}, it is quite possible that together all these methods will allow to push the coating thermal noise below the quantum noise.

Concerning the quantum noise, it has to be emphasized, that only two simplest configurations which do not use the advanced techiques like optical springs \cite{99a1BrKh, Buonanno2001}, squeezed light \cite{Caves1981}, filter cavities \cite{02a1KiLeMaThVy, Corbitt2004-3} {\it etc} were considered here. All these methods of shaping the quantum noise, developed for the Michelson/Fabry-Perot topology, can be used in the Mach-Zehnder/Fabry-Perot case as well, with the additional flexibility provided by the two output ports.

\acknowledgements

This work has been supported by Russian Foundation for Basic Research Grant No. 08-02-00580-a and NSF and Caltech Grant No. PHY-0651036. The author is grateful to Stefan Danilishin and Rana Adhikari for the useful remarks and suggestions.

\appendix

\section{Notations and approximations}

High (optical range) frequencies are denoted by $\omega$, and low (mechanical-range) ones by $\Omega$. Typically, $\omega=\omega_p+\Omega$. It is supposed that the following inequalities are fulfilled:
\begin{equation}
  \Omega, \gamma \ll \frac{L}{\tau} \ll \omega_o \,.
\end{equation} 

The field amplitudes are presented as sums of large classical values (denoted by capital roman letters) and small quantum ones (denoted by small roman letter). Only linear in these small quantum fluctuations and in the mirrors displacements terms are kept. Two-photon quadrature vectors \cite{Caves1985, Schumaker1985} are denoted by bold-face roman letters:
\begin{subequations}
  \begin{gather}
    \hat{\bf a}(\Omega) = \svector{\hat{\rm a}^c(\Omega)}{\hat{\rm a}^s(\Omega)} \,, \\
    \hat{\rm a}^c(\Omega) = \frac{
        \hat{\rm a}(\omega_o+\Omega) + \hat{\rm a}^+(\omega_o-\Omega)
      }{\sqrt{2}}\,,\\
    \hat{\rm a}^s(\Omega) = \frac{
        \hat{\rm a}(\omega_o+\Omega) - \hat{\rm a}^+(\omega_o-\Omega)
      }{i\sqrt{2}}\,.
  \end{gather}
\end{subequations}
In vacuum state, single-sided spectral densities of the quadrature amplitudes are equal to 1.

In most cases, dependence of the field operators, as well as of the mechanical positions and forces spectra, on the frequencies $\omega$, $\Omega$ is omitted below for brevity.

The ``down'' and ``up'' arms of the interferometers are denoted by the subscript $a=d,u$. 

All optical losses are modeled here by two grey filters with the power transmittances $\eta$, placed before the detectors:
\begin{align}\label{loss}
  & \hat{\rm q}_{1,2} = \sqrt{\eta}\,\hat{\rm b}_{1,2\,-}
    + \sqrt{1-\eta}\,\hat{\rm n}_{1,2} \,,
\end{align}
where $\hat{\rm n}_{1,2}$ are the vacuum noises associated with these losses.

The beamplitters are described by the following equations:
\begin{subequations}\label{bs}
  \begin{gather}
    \svector{\hat{\rm a}_{1,2\,d}}{\hat{\rm a}_{1,2\,u}}
      = \frac{1}{\sqrt{2}}\smatrix{1}{1}{1}{-1}
        \svector{\hat{\rm a}_{1,2\,+}}{\hat{\rm a}_{1,2\,-}} , \\
    \svector{\hat{\rm b}_{1,2\,+}}{\hat{\rm b}_{1,2\,-}}
    = \frac{1}{\sqrt{2}}\smatrix{1}{1}{1}{-1}
      \svector{\hat{\rm b}_{1,2\,d}}{\hat{\rm b}_{1,2\,u}} .
  \end{gather}
\end{subequations}

Some other notations used in the Appendices and not shown in Table\,\ref{tab:notations} are listed below:
\begin{gather}
  s = -i\Omega \,, \\
  \mathbb{Y} = \smatrix{0}{-1}{1}{0} , \\
  \alpha = \dfrac{2\omega_o}{\sqrt{\gamma cL}} \,.
\end{gather}

\section{Quantum noise of the two-port position meter}\label{app:pm}

\begin{figure}
  \includegraphics{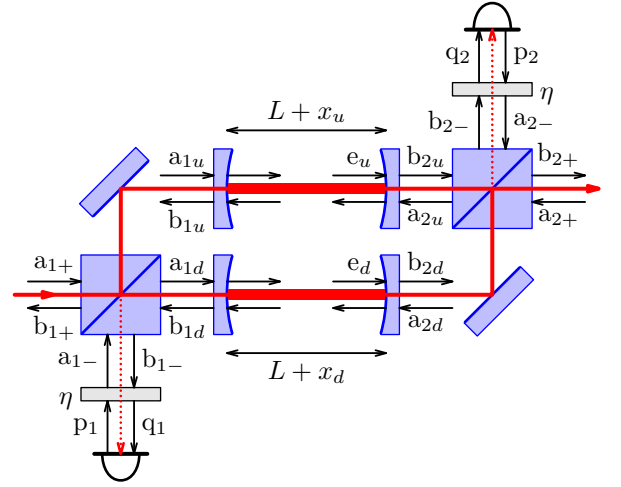}
  \caption{To calculation of the Mach-Zehnder/Fabry-Perot interferometer quantum noises.}\label{fig:mach_zehnder_calc}
\end{figure} 

A more detailed scheme of the Mach-Zehnder/Fabry-Perot interferometer is shown in Fig.\,\ref{fig:mach_zehnder_calc}.
Input-output relations for the Fabry-Perot cavities are derived, in particular, in paper \cite{06a2Kh}. In the case of the identical input and end mirrors, they have the following form:
\begin{subequations}\label{pm_fp}
  \begin{gather}
    \hat{\rm b}_{1,2\,a} = \frac{
        -s\hat{\rm a}_{1,2\,a} - \gamma\hat{\rm a}_{2,1\,a}
        \pm \alpha\gamma{\rm E}x_a/\sqrt{2}
      }{\gamma+s} \,, \\
    \hat{\rm e}_a = i\sqrt{\frac{\gamma}{2\tau}}\,\frac{
        \hat{\rm a}_{1a} - \hat{\rm a}_{2a} + \alpha{\rm E}x_a/\sqrt{2}
      }{\gamma+s} \,,
    \end{gather}
\end{subequations}
where $x_a$ are variations of the Fabry-Perot cavities lengths.

Combining Eqs.\,(\ref{loss}, \ref{bs}, \ref{pm_fp}), and switching to the quadrature amplitudes notations, we obtain, that:
\begin{subequations}\label{pm_det}
  \begin{gather}
    \hat{\bf q}_{1,2} = \pm\sqrt{\eta}
      \biggl(\hat{\bf j}_{1,2} + \frac{\alpha\gamma{\bf E}x}{\gamma+s}\biggr)
      + \sqrt{1-\eta}\,\hat{\bf n}_{1,2} , \\
    \hat{\bf e} = \frac{{\bf e}_d-{\bf e}_u}{\sqrt{2}}
      = \mathbb{Y}\sqrt{\frac{\gamma}{2\tau}}\biggl(
        \frac{\hat{\bf j}_1 + \hat{\bf j}_2}{\gamma-s} + \frac{\alpha{\bf E}x}{\gamma+s}
      \biggr) , \label{pm_ce}
  \end{gather}
\end{subequations}
where
\begin{gather}\label{j_pm}
  {\bf E} = \svector{\sqrt{2}{\rm E}}{0} , \\
  x = \frac{x_1-x_2}{2} \,, \\
  \hat{\bf j}_{1,2}
    = \mp\frac{s\hat{\bf a}_{1,2\,-} + \gamma\hat{\bf a}_{2,1\,-}}{\gamma+s} \,.
\end{gather} 

The photocurrents are proportional to
\begin{equation}
  i_{1,2} \propto \Phi^{\sf T}_{1,2}\hat{\bf q}_{1,2}
    \propto \hat{x}_{1,2}^{\rm fl} + x \,,
\end{equation} 
where
\begin{equation}
  \Phi_{1,2} = \svector{\cos\phi_{1,2}}{-\sin\phi_{1,2}} ,
\end{equation}
$\phi_{1,2}$ are the homodyne angles, and
\begin{equation}\label{pm_x_fl} 
  \hat{x}_{1,2}^{\rm fl}
  = \frac{\gamma+s}{\sqrt{2}\alpha\gamma{\rm E}\cos\phi_{1,2}}\,
    \Phi^+_{1,2}
      \biggl(\hat{\bf j}_{1,2} \pm \sqrt{\frac{1-\eta}{\eta}}\,\hat{\bf n}_{1,2}\biggr)
\end{equation}
are the measurement noises.

The back-action forces, acting on the cavities mirrors, are equal to
\begin{equation}\label{F_12_fl}
  \hat{F}_a^{\rm fl} = \frac{2\hbar\omega_o{\bf E}^+\hat{\bf e}_a}{c}\,.
\end{equation}
The differential force is equal to [see Eq.\,\eqref{pm_ce}]
\begin{equation}
  \hat{F}^{\rm fl} = \hat{F}_d^{\rm fl} - \hat{F}_u^{\rm fl}
  = \frac{2\sqrt{2}\hbar\omega_o{\bf E}^+\hat{\bf e}}{c}
  = \hat{F}_1^{\rm fl} + \hat{F}_2^{\rm fl} \,,
\end{equation}
where
\begin{equation}\label{pm_F_fl}
  \hat{F}_{1,2}^{\rm fl}
  = -\frac{\sqrt{2}\hbar\alpha\gamma{\rm E}\hat{\rm j}_{1,2}^s}{\gamma-s} \,.
\end{equation}

Suppose that $\hat{\bf a}_{1,2\,-}$ correspond to two non-correlated vacuum noises. It is easy to show that in this case $\hat{\bf j}_{1,2}$ also correspond to (some other) two  non-correlated vacuum noises. Therefore, two detectors outputs corresponds to two independent meters with the noises defined by  $\hat{\bf j}_{1,2}$. It follows from Eqs.\,(\ref{pm_x_fl}, \ref{pm_F_fl}), that the noise spectral densities of these meters are equal to
\begin{subequations}\label{pm_SxSF}
  \begin{gather}
    S_{1,2}^x
      = \frac{\hbar cL(\gamma^2+\Omega^2)}{8\omega_oI_c\gamma\eta\cos^2\phi_{1,2}} \,, \\
    S_{1,2}^F = \frac{8\hbar\omega_oI_c\gamma}{cL(\gamma^2+\Omega^2)} \,, \\
    S_{1,2}^{xF} = \hbar\tan\phi_{1,2} \,.
  \end{gather}
\end{subequations}

\section{Quantum noise of two-port interferometer with central mirrors}\label{app:sm}

\subsection{Arm cavity}


\subsubsection{Field amplitudes}

\begin{figure}
  \includegraphics{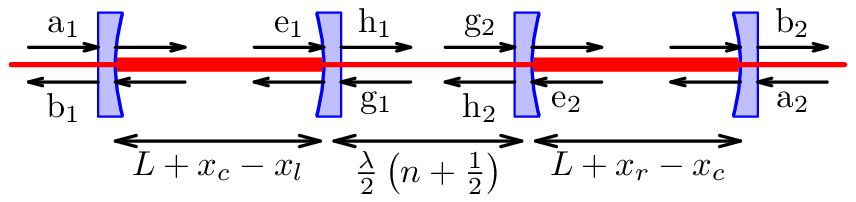} \\
  \includegraphics{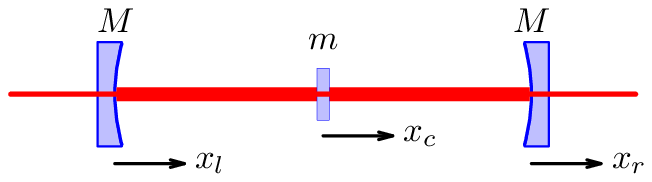}
  \caption{One arm of the two-port speed meter: top --- optical scheme, bottom --- mechanical scheme.}\label{fig:speedm_arm}
\end{figure} 

Each of the arm cavities of the speed meter interferometer shown in Fig.\,\ref{fig:speedmeter} can be presented as a sequence of two identical Fabry-Perot cavities, see Fig.\,\ref{fig:speedm_arm}, top. Therefore, it can be decribed by two sets of equations describing each of these Fabry-Perot cavities:
\begin{subequations}\label{speedm_bhe} 
  \begin{gather}
    \hat{\rm b}_{1,2\,a} = \frac{
        -s\hat{\rm a}_{1,2\,a} - \gamma\hat{\rm g}_{1,2\,a}
        + \alpha\gamma{\rm E}_{1,2}x_{1,2\,a}/\sqrt{2}
      }{\gamma+s} \,, \\
    \hat{\rm h}_{1,2\,a} = \frac{
        -\gamma\hat{\rm a}_{1,2\,a} - s\hat{\rm g}_{1,2\,a}
        - \alpha\gamma{\rm E}_{1,2}x_{1,2\,a}/\sqrt{2}
      }{\gamma+s} \,, \\
    \hat{\rm e}_{1,2\,a}
    = i\sqrt{\frac{\gamma}{2\tau}}\,\frac{
        \hat{\rm a}_{1,2\,a} - \hat{\rm g}_{1,2\,a}
        + \alpha{\rm E}_{1,2}x_{1,2\,a}/\sqrt{2}
      }{\gamma+s} \,,
  \end{gather}
\end{subequations}
[compare with Eqs.\,\eqref{pm_fp}], and additional equations describing the coupling between them:
\begin{equation}
  \hat{\rm g}_{1,2\,a} = e^{i\theta}\hat{\rm h}_{2,1\,a} \,. \label{speedm_gh}
\end{equation}
Here
\begin{align}
  & x_{1a} = x_{ca}-x_{la} \,, & & x_{2a} = x_{ra}-x_{ca} \,,
\end{align}
\begin{equation}
  {\rm E}_2 = e^{i\theta}({\rm E}_1 \equiv {\rm E}) \,,
\end{equation}
and
\begin{equation}
  \theta = \frac{\pi}{2} + \pi n
\end{equation} 
is the phase shift introduced by the central mirror. Note that while the intermediate equations depend on whether $n$ is even or odd, the final ones do not. Odd $n$ will be used in the calculations below, giving
\begin{equation}
  e^{i\theta} = -i \,.
\end{equation}
Solution of these equations is the following:
\begin{subequations}\label{sm_arm} 
  \begin{gather}
    \hat{\rm b}_{1,2\,a} = \frac{
        -2s^2\hat{\rm a}_{1,2\,a} - i\gamma^2\hat{\rm a}_{2,1\,a}
        \pm \sqrt{2}\,\alpha\gamma{\rm E}_{1,2}v_{1,2\,a}
      }{\gamma^2 + 2\gamma s + 2s^2} \,, \\
    \hat{\rm e}_{1,2\,a} = i\sqrt{\frac{\gamma}{2\tau}}\,\frac{
        (\gamma+2s)\hat{\rm a}_{1,2\,a} - i\gamma\hat{\rm a}_{2,1\,a}
        \pm \sqrt{2}\,\alpha{\rm E}_{1,2}v_{1,2\,a}
      }{\gamma^2 + 2\gamma s + 2s^2} \,,
  \end{gather}
\end{subequations}
where
\begin{equation}
  v_{1,2\,a} = X_{1,2}x_a + Y_{1,2}y_a  \,,
\end{equation} 
\begin{subequations}
  \begin{gather}
    x_a = \frac{x_{1a} - x_{2a}}{2} = \frac{2x_{ca} - x_{ra} - x_{la}}{2} \,, \\
    y_a = \frac{x_{1a} + x_{2a}}{2} = \frac{x_{ra} - x_{la}}{2} 
  \end{gather}
\end{subequations}
[see also Eqs.\,\eqref{sm_XY}].

In two-photons quadratures notations, Eqs.\,\eqref{sm_arm} has the following form:
\begin{subequations}\label{sm_arm_q} 
  \begin{gather}
    \hat{\bf b}_{1,2\,a} = \frac{
        -2s^2\hat{\bf a}_{1,2\,a} - \mathbb{Y}\gamma^2\hat{\bf a}_{2,1\,a}
        \pm \sqrt{2}\,\alpha\gamma{\bf E}_{1,2}v_{1,2\,a}
      }{\gamma^2 + 2\gamma s + 2s^2} \,, \label{sm_arm_q(a)}  \\
    \hat{\bf e}_{1,2\,a} = \mathbb{Y}\sqrt{\frac{\gamma}{2\tau}}\,\frac{
        (\gamma+2s)\hat{\bf a}_{1,2\,a} - \mathbb{Y}\gamma\hat{\bf a}_{2,1\,a}
        \pm \sqrt{2}\,\alpha{\bf E}_{1,2}v_{1,2\,a}
      }{\gamma^2 + 2\gamma s + 2s^2} \,,
  \end{gather}
\end{subequations}
where
\begin{equation}
  {\bf E}_2 = -\mathbb{Y}
  \left[{\bf E_1} \equiv {\bf E} = \svector{\sqrt{2}{\rm E}}{0}\right] .
\end{equation} 

\subsubsection{Radiation-pressure forces}

Mechanical equations of motion for the mirrors are the following:
\begin{subequations}
  \begin{gather}
    Ms^2\hat{x}_{la} = -\hat{F}_{1a} \,, \\
    ms^2\hat{x}_{ca} = \hat{F}_{1a} - \hat{F}_{2a} \,, \\
    Ms^2\hat{x}_{ra} = \hat{F}_{2a} \,,
  \end{gather}
\end{subequations}
where
\begin{equation}
  \hat{F}_{1,2\,a} = \frac{2\hbar\omega_o{\bf E}_{1,2}\hat{\bf e}_{1,2\,a}}{c}
\end{equation}
are the the radiation-pressure forces inside the left and the right cavities. Therefore,
\begin{align}\label{app_sm_eqs}
  & \hat{x}_a = \frac{\hat{F}_{xa}}{\mu s^2} \,, &
  & \hat{y}_a = \frac{\hat{F}_{ya}}{2M s^2} \,,
\end{align}
where
\begin{subequations}
  \begin{gather}
    \hat{F}_{xa} = \hat{F}_{1a} - \hat{F}_{2a}
      = \frac{2\hbar\omega_o{\bf E}\hat{\bf e}_{xa}}{c} \,, \\
    \hat{F}_{ya} = \hat{F}_{1a} + \hat{F}_{2a}
      = \frac{2\hbar\omega_o{\bf E}\hat{\bf e}_{ya}}{c} \,,
  \end{gather}
\end{subequations}
and
\begin{subequations}
  \begin{gather}
    \hat{\bf e}_{xa} = \sqrt{\frac{2\gamma}{\tau}}\,
      \frac{s\mathbb{Y}\hat{\bf a}_{1a} + (\gamma+s)\hat{\bf a}_{2a}}
        {\gamma^2 + 2\gamma s + 2s^2} \,, \\
    \hat{\bf e}_{ya} = \sqrt{\frac{2\gamma}{\tau}}\,
      \frac{(\gamma+s)\mathbb{Y}\hat{\bf a}_{1a} - s\hat{\bf a}_{2a}}
        {\gamma^2 + 2\gamma s + 2s^2} \,.
  \end{gather}
\end{subequations}
[see also Eq.\,\eqref{sm_mu}]

\subsection{The noises spectral densities}

\begin{figure}
  \includegraphics{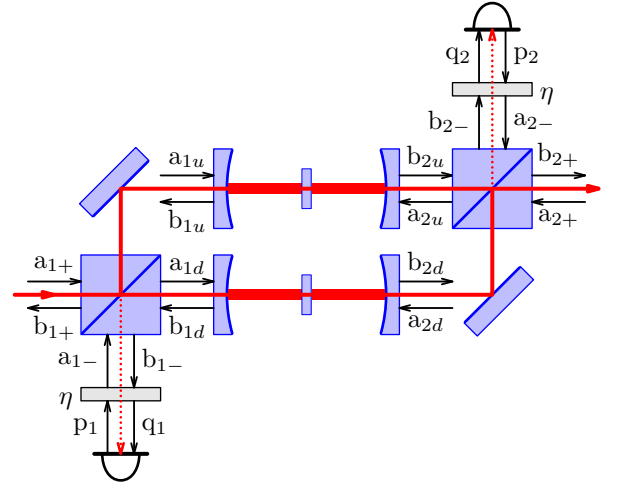}
  \caption{To calculation of the Mach-Zehnder/Fabry-Perot speed meter quantum noises.}\label{fig:speedmeter_calc}
\end{figure} 

Return to the full two-arm scheme shown in Fig.\,\ref{fig:speedmeter_calc}. Combining Eqs.\,\eqref{sm_arm_q} for each of the arms with Eqs.\,(\ref{loss}, \ref{bs}), we obtain, that
\begin{multline}
  \hat{\bf q}_{1,2} = \sqrt{\eta}\,\frac{
      -2s^2\hat{\bf a}_{1,2\,-} - \mathbb{Y}\gamma^2\hat{\bf a}_{2,1\,-}
      \pm 2\alpha\gamma{\bf E}_{1,2}v_{1,2}
    }{\gamma^2 + 2\gamma s + 2s^2} \\
    + \sqrt{1-\eta}\,\hat{\bf n}_{1,2} \,, 
\end{multline}
where
\begin{equation}
  v_{1,2} = X_{1,2}x + Y_{1,2}y \,,
\end{equation} 
\begin{align}
  & x = \frac{x_d-x_u}{2} \,, & & y = \frac{y_d-y_u}{2} \,,
\end{align}
[see also Eqs.\,\eqref{sm_XY}], and the differential radiation-pressure forces are equal to
\begin{equation}\label{sm_F_xy} 
  \hat{F}_{x,y}^{\rm fl} = \hat{F}_{u\,x,y} - \hat{F}_{d\,x,y}
    = \frac{2\sqrt{2}\hbar\omega_o{\bf E}\hat{\bf e}_{x,y}}{c} \,,
\end{equation} 
where
\begin{subequations}
  \begin{gather}
    \hat{\bf e}_x = \frac{\hat{\bf e}_{dx}-\hat{\bf e}_ux}{\sqrt{2}}
      =\sqrt{\frac{2\gamma}{\tau}}\,
        \frac{s\mathbb{Y}\hat{\bf a}_{1-} + (\gamma+s)\hat{\bf a}_{2-}}
          {\gamma^2 + 2\gamma s + 2s^2} \,, \\
    \hat{\bf e}_y = \frac{\hat{\bf e}_{dy}-\hat{\bf e}_uy}{\sqrt{2}}
      = \sqrt{\frac{2\gamma}{\tau}}\,
      \frac{(\gamma+s)\mathbb{Y}\hat{\bf a}_{1-} - s\hat{\bf a}_{2-}}
        {\gamma^2 + 2\gamma s + 2s^2} \,.
  \end{gather}
\end{subequations}

Introduce two new noises:
\begin{equation}\label{sm_j} 
  \hat{\bf j}_{1,2} = \frac{
      -2s^2\hat{\bf a}_{1,2\,-} - \mathbb{Y}\gamma^2\hat{\bf a}_{2,1\,-}
    }{\gamma^2 + 2\gamma s + 2s^2} \,.
\end{equation}
In these notations,
\begin{equation}
  \hat{\bf q}_{1,2} = \sqrt{\eta}\,\left(
    \hat{\bf j}_{1,2}
    \pm \frac{2\alpha\gamma{\bf E}_{1,2}v_{1,2}}{\gamma^2 + 2\gamma s + 2s^2}
  \right) + \sqrt{1-\eta}\,\hat{\bf n}_{1,2} \,, 
\end{equation} 
\begin{subequations}
  \begin{gather}
    \hat{\bf e}_x = \sqrt{\frac{2\gamma}{\tau}}\,
      \frac{X_1^*\mathbb{Y}\hat{\bf j}_1 + X_2^*\hat{\bf j}_2}
          {\gamma^2 - 2\gamma s + 2s^2} \,, \\
    \hat{\bf e}_y = \sqrt{\frac{2\gamma}{\tau}}\,
      \frac{Y_1^*\mathbb{Y}\hat{\bf j}_1 + Y_2^*\hat{\bf j}_2}
        {\gamma^2 - 2\gamma s + 2s^2} \,,
  \end{gather}
\end{subequations}

The photocurrents are proportional to
\begin{equation}\label{sm_i_12} 
  i_{1,2} \propto \Phi_{1,2}^{\sf T}\hat{\bf q}_{1,2}
    \propto v_{1,2} + \hat{v}_{1,2}^{\rm fl} \,,
\end{equation}
where
\begin{align}
  & \Phi_1 = \svector{\cos\phi_1}{-\sin\phi_2} \,, &
  & \Phi_2 = \svector{\sin\phi_2}{\cos\phi_2}
\end{align}
(the second homodyne angle is redefined as $\phi+\pi/2\to\phi$ for the notations consistency),
\begin{equation}\label{sm_v_fl_12}
  v_{1,2}^{\rm fl}
  = \frac{\gamma^2 + 2\gamma s + s^2}{2\sqrt{2}\,\alpha\gamma{\rm E}\cos\phi_{1,2}}\,
    \Phi_{1,2}^{\sf T}
    \left[\hat{\bf j}_{1,2} + \sqrt{\frac{1-\eta}{\eta}}\,\hat{\bf n}_{1,2}\right] .
\end{equation} 

Using the noises \eqref{sm_j}, the back-action forces \eqref{sm_F_xy} can be rewritten in the form \eqref{sm_F_fl_xy}, with
\begin{subequations}
  \begin{gather}
    \hat{p}_1^{\rm fl}
      = -\frac{2\sqrt{2}\hbar\alpha\gamma{\rm E}}{\gamma^2 - 2\gamma s + 2s^2}\,
        \hat{\rm j}_1^s  \,, \\
    \hat{p}_2^{\rm fl}
      = \frac{2\sqrt{2}\hbar\alpha\gamma{\rm E}}{\gamma^2 - 2\gamma s + 2s^2}\,
        \hat{\rm j}_2^c \,.
  \end{gather}
\end{subequations}

Similar to Eq.\,\eqref{j_pm}, if $\hat{\bf a}_{1,2\,-}$, correspond to two non-correlated vacuum noises. then  $\hat{\bf j}_{1,2}$ also correspond to two  non-correlated vacuum noises. In this case, spectral densities of the noises $\hat{v}_{1,2}$ and $\hat{p}_{1,2}$ are equal to
\begin{subequations}\label{sm_S_vp} 
  \begin{gather}
    S^v_{1,2} = \frac{\hbar}{\mu\mathcal{K}\eta\cos^2\phi_{1,2}} \,, \\
    S^p_{1,2} = \hbar\mu\mathcal{K} \,, \\
    S_{vp}^{1,2} = \hbar\tan\phi_{1,2} \,.
  \end{gather}
\end{subequations}
[see also Eqs.\,\eqref{sm_varK}].

\subsection{Quantum noise of the two-port speed meter}\label{app:sm_S_sum}

If $\phi_1=\pi/2$, then the output signals of the first (auxiliary) and the second (main) homodyne detector output are proportional to [see Eqs.\,(\ref{sm_i_12}, \ref{sm_v_fl_12})]:
\begin{subequations}
  \begin{gather}
    i_1 \propto a = \frac{\hat{v}_1^{\rm fl}}{\tan\phi_1}\bigg|_{\phi_1\to\pi/2}
      \label{sm_a} , \\
    i_2 \propto x + \frac{Y_2y + v_2^{\rm fl}}{X_2}
      = \hat{x}_0 + \hat{x}^{\rm fl} + \frac{\hat{F}^{\rm fl}_{\rm raw}}{\mu s^2} \,,
  \end{gather}
\end{subequations}
where $\hat{x}_0(\Omega)$ of the intrinsic motion of the test object,
\begin{equation}\label{sm_x} 
  \hat{x}^{\rm fl} = \frac{v_2^{\rm fl}}{X_2}
\end{equation} 
is the measurement noise, recalculated as the equivalent fluctuational displacement of the ``$x$'' degree of freedom, 
\begin{equation}\label{sm_F_raw} 
  \hat{F}^{\rm fl}_{\rm raw}
  = \hat{F}_x^{\rm fl} + \kappa\frac{Y_2}{X_2}\,\hat{F}_y^{\rm fl}
  = G_1\hat{p}_1^{\rm fl} + i\Omega G_2\hat{p}_2^{\rm fl} \,,
\end{equation} 
is the ``raw'' back action noise (it takes into account perturbation of the ``$y$'' degree of freedom, but not the information provided by the auxiliary detector),
\begin{equation}
  \kappa = \frac{\mu}{2M} \,,
\end{equation}
and
\begin{subequations}
  \begin{gather}
    G_1 = \gamma(1+\kappa) + i\Omega(1-\kappa) \,, \\
    G_2 = 1 + \kappa\frac{\gamma^2+\Omega^2}{\Omega^2} \,.
  \end{gather}
\end{subequations}

It follows from Eqs.\,\eqref{sm_S_vp}, that spectral densities of the noises (\ref{sm_a}, \ref{sm_x}, and \ref{sm_F_raw}) are equal to
\begin{subequations}\label{sm_S_x} 
  \begin{gather}
    S^x = \frac{\hbar}{\mu\Omega^2\mathcal{K}\cos^2\phi_2} \,, \\
    S^{xF} = \hbar G_2\tan\phi_2 \,,
  \end{gather}
\end{subequations}
\begin{subequations}
  \begin{gather}
    S^a = \frac{\hbar}{\mu\mathcal{K}\eta} \,, \\
    S^{aF} = \hbar G_1 \,, \\
    S^F_{\rm raw} = \hbar\mu\mathcal{K}\left(|G_1|^2 + \Omega^2G_2^2\right) .
  \end{gather}
\end{subequations}
Note that the auxiliary output $a$ is correlated with the back raw action force $F_{\rm raw}^{\rm fl}$, but is not correlated with the measurement noise $x^{\rm fl}$. Therefore, the optimal date processing \eqref{x12_to_y} is this case will affect only the effective back action noise, giving the following spectral density:
\begin{equation}\label{sm_S_F} 
  S^F_{\rm eff} = S^F_{\rm raw} - \frac{|S^{aF}|^2}{S^a}
  = \hbar\mu\mathcal{K}\left[(1-\eta)|G_1|^2 + \Omega^2|G_2|^2\right] .
\end{equation} 
The corresponding sum noise spectral density is equal to
\begin{multline}\label{sm_S_sum_k}
  S_{\rm sum} = S^x - \frac{2S^{xF}}{\mu\Omega^2} + \frac{S^F_{\rm eff}}{\mu^2\Omega^4}\\
  = \frac{\hbar}{\mu\Omega^2}\biggl\{
      \frac{1}{\mathcal{K}\eta\cos^2\phi_2} - 2G_2\tan\phi_2 \\
      + \mathcal{K}\left[\frac{(1-\eta)|G_1|^2}{\Omega^2} + G_2^2\right]
    \biggr\} .
\end{multline}

If $\kappa\ll1$, then the homodyne angle
\begin{equation}
  \phi_2 = \mathcal{K}_0(1+\kappa)\eta
\end{equation} 
provides the sum noise minimum of in the broad frequency band $\gamma\sqrt{\kappa}\ll\Omega\ll\gamma$. In this case,
\begin{multline}\label{sm_S_sum_k_opt}
  S_{\rm sum} = \frac{\hbar}{\mu\Omega^2}\Biggl(
      \frac{1}{\mathcal{K}\eta} 
      + \mathcal{K}\eta\biggl\{
          \left[
            (1+\kappa)\frac{4\Omega^4}{\gamma^4} - \kappa\frac{\gamma^2}{\Omega^2}
          \right]^2 \\
          + \frac{1-\eta}{\eta}\left[\frac{|G_1|^2}{\Omega^2} + G_2^2\right]
        \biggr\}
    \Biggr) .
\end{multline}


\end{document}